\documentstyle[11pt,fleqn]{article}

\def\beeq{\begin{equation}}
\def\eneq{\end{equation}}
\def\beeqa{\begin{eqnarray}}
\def\eneqa{\end{eqnarray}}

\def\soc{{\rm C}_{60}}

\addtocounter{section}{-1}
\begin{document}

\begin{center}

{\large{\bf{Dispersion of the Third-Order\\
Nonlinear Optical Susceptibility in C$_{60}$\\
Calculated with a Tight-Binding Model
} } }

\vspace{1cm}

{\rm Kikuo H{\sc arigaya}\footnote[1]{e-mail: harigaya@etl.go.jp,
e9118@jpnaist.bitnet} and Shuji A{\sc be}}\\

\vspace{1cm}

{\sl Fundamental Physics Section, Physical Science Division,\\
Electrotechnical Laboratory,\\
Umezono 1-1-4, Tsukuba, Ibaraki 305}

\vspace{1cm}

(Received April 3, 1992; accepted for publication May 16, 1992)
\end{center}

\Roman{table}

\vspace{1cm}

\noindent
The frequency dependence of third harmonic generation (THG) in C$_{60}$
is calculated, making use of a tight-binding model for $\pi$-electrons.
The magnitudes of the THG, about 10$^{-12}$ esu, near zero frequency,
agree with those in experiments for the low-energy region.
We can also explain the order of the magnitude, 10$^{-11}$esu, around the
three-photon resonance peak due to the lowest allowed excitation,
recently measured by Meth et al.  At higher energies, we predict a
large enhancement of the THG at $3\omega \sim 6$eV as a result of double
resonance enhancement.

{}~~~~~~~


\pagebreak

\section{Introduction}

Recently, the buckminsterfullerene, $\soc$, has received much attention
and has been intensively investigated.  The $\soc$ molecule is shaped
like a soccerball.  Besides the structural uniqueness, the $\soc$
molecule and solids show various interesting properties:
superconductivity,$^{1)}$ ferromagnetism,$^{2)}$ optical
nonlinearity,$^{3-4)}$ and so on.

The recent paper by Blau et al.$^{3)}$ has reported that the magnitude
of the nonlinear susceptibility per $\soc$ molecule is $|\gamma|
= 1.5 \times 10^{-42}$m$^5 \cdot$V$^{-2}$, which is the same size as
observed in polydiacetylene.  Also, the paper by Hoshi et al.$^{4)}$
has shown the relatively large third harmonic generation (THG),
$|\chi^{(3)} (3\omega;\omega,\omega,\omega)| \simeq 2 \times 10^{-10}$esu,
of a $\soc$ film.
Therefore, the large optical nonlinearity of $\soc$ is quite attractive
in view of its technological importance as well as scientific interest.
Meth et al.\footnote[1]{J. S. Meth, H. Vanherzeele and Y. Wang: preprint.}
 have measured the dispersion
(the dependence on the photon energy) of the THG.  They have
concluded that there is a three-photon resonance at $\omega = 0.94$eV,
whose third harmonic energy corresponds to the dipole allowed transition
at 2.8eV in the solid $\soc$.  The magnitude of the THG is
$4.1 \times 10^{-12}$esu away from the resonance, while it is
$2.7 \times 10^{-11}$esu at the peak.  The magnitude by their
measurement is smaller than that in ref. 4 by the factor $10^{-2}
- 10^{-1}$.  The intrinsic value of $\soc$ itself has not been
clarified yet.  This might be due to different experimental
circumstances.

In this paper, we calculate the dispersion of the THG in order to
obtain knowledge of the possible spectral shape.  In addition,
we would like to know the fundamental magnitude of the THG,
calculated from the viewpoint of
a microscopic theory.  This information is helpful
when we analyze experimental data.  We use a simple tight-binding model
where $\pi$-electrons on the $\soc$ ball are considered.
In the model, the strong hopping integral is assigned to the
double bonds, and the weak one to the single bonds.  We do not
include Coulomb interactions among electrons, because this paper
is at the first stage of the investigation of nonlinear optics of
$\soc$.  The intrinsic magnitudes of the THG spectra will be
determined at least from dipole interactions given by the one-electron picture.
The Coulomb interactions would be effective as perturbations.  We use
the general formula shown in refs. 5-7.

The main conclusions are as follows: (1) The magnitudes of the THG,
about 10$^{-12}$ esu, at frequency zero, and the order 10$^{-11}$esu
at the peak of the lowest allowed transition, agree well with those
in  experiments.$^*$\footnote[2]{Z. H. Kafafi, J. R. Lindle, R. G. S. Pong,
F. J. Bartoli, L. J. Lingg, and J. Milliken: preprint.}
(2) The remarkably large peaks can be
measured at three-photon energies, 6.1 and 6.3eV, owing to the
double resonance enhancement.

In the next section, the tight-binding model is given and the parameters
are explained.  The formula of the THG is also shown.  In \S 3, we report
results.  We close the paper with several remarks in \S 4.

\section{Model and Formulation}

The tight-binding model hamiltonian for $\soc$ is:
\beeq
\Ham = {\sum_{\langle i,j \rangle, \sigma}}^{\rm D} (-t_{\rm D})
(c_{i,\sigma}^\dagger c_{j,\sigma} + {\rm h.c.})
+ {\sum_{\langle i,j \rangle,\sigma}}^{\rm S} (-t_{\rm S})
(c_{i,\sigma}^\dagger c_{j,\sigma} + {\rm h.c.}),
\eneq
where $c_{i,\sigma}$ is an annihilation operator of a $\pi$-electron
with spin $\sigma$; the sum with the symbol D (or S) is taken over for the
pairs of neighboring atoms, $\langle i,j \rangle$, along the double (or single)
bond; and $t_{\rm D}$ and $t_{\rm S}$ are the hopping integrals.
In this paper, we use the parameters $t_{\rm D} = 2.337$eV and
$t_{\rm S} = 1.982$eV.  These give the total energy width of the
$\pi$-electron states, 11.84eV, and the energy gap, 2.004eV.
The magnitudes are similar to those in the literature, refs. 8-12.

We assume in this paper that Coulomb interaction is not so important.
The effective one-electron orbitals can describe the energy
level structures well, as shown by the optical absorption.$^{8)}$
Also, interactions between $\soc$ balls are not considered,
because the hopping interaction strength can be estimated to be of the
order 0.1eV at most; this does not smear out the ordering
of the electronic levels of the molecule.$^{9)}$

The THG is calculated with the conventional formula:$^{5-7)}$
\beeqa
&~&\chi^{(3)} (3\omega;\omega,\omega,\omega) = e^4 N
\sum_{n,m,l} f_{g,n} f_{n,m} f_{m,l} f_{l,g} \nonumber \\
&\times& [ \frac{1}{(E_{n,g}-3\omega)(E_{m,g}-2\omega)(E_{l,g}-\omega)}
+ \frac{1}{(E_{n,g}^*+\omega)(E_{m,g}-2\omega)(E_{l,g}-\omega)} \nonumber \\
&+& \frac{1}{(E_{n,g}^*+\omega)(E_{m,g}^*+2\omega)(E_{l,g}-\omega)}
+ \frac{1}{(E_{n,g}^*+\omega)(E_{m,g}^*+2\omega)(E_{l,g}^*+3\omega)} ],
\eneqa
where $N=1.397\times 10^{21} {\rm cm}^{-3}$ is the number density of
the $\soc$ solid,
$f_{m,l}$ is the dipole matrix element, and $E_{n,g} = E_n - E_g$.
Here, $E_n$ is the energy of the excited state and $E_g$ is the energy
of the ground state.  In the actual calculation, we change the order
of terms so as to take into account the mutual cancellation among
them.$^{13)}$  Also, we include a small imaginary part in the denominator:
for example, $E_{n,g} \rightarrow E_{n,g} + {\rm i} \eta$ and
$E_{n,g}^* \rightarrow E_{n,g}^* - {\rm i} \eta$.
This assumes a lifetime broadening, and suppresses the height of
the delta-function peaks.      We make the value of
$\eta$ as small as possible, in order to see the structures
of the dispersions clearly, though the realistic value would be much
larger.  In the next section, we report the results with the value
$\eta = 1.68 \times 10^{-2}$eV.

\section{Dispersion of Optical Spectra}

First, we show the spectrum of single-photon absorption calculated with a
formula similar to eq. (2), for the sake of checking the calculation.
The result is depicted in Fig. 1.  In the lower part of the figure, the
energy values of the dipole allowed and
forbidden excitations are shown.  The electronic states in the energy
region between 0eV and 8eV mainly come from $\pi$-orbitals.
Therefore, the present calculation by the simple tight-binding model
would be reliable enough for this energy region.
The transition between the highest occupied molecular orbital (HOMO) and
the lowest unoccupied molecular orbital (LUMO) is forbidden.  This accords
with the previous calculations.$^{8,11,12)}$
The peak with the lowest energy corresponds to the transition between the
HOMO and the next LUMO.  The peak with the next lowest energy is due
to the transition between the next HOMO and the LUMO.
In the experiments (Meth et al. and ref. 14), these two peaks merge
into a single peak due to larger broadening.
Other features of the higher energy region agree well with
the previous independent calculation (see Fig. 2(a) of ref. 8).

Next, we look at spectral dispersions of the THG.  We display the calculated
results of the real and imaginary parts, and the absolute value, in Fig. 2.
In the bottom of the figure,
we show the energies of the dipole allowed excitations, where three-photon
resonances can appear, and also the energies of the forbidden excitations
multiplied by $3/2$, where two-photon resonances can appear.  The
peaks in the THG spectrum can be assigned as two- or three-photon
resonances.   Major peaks can be assigned as three-photon
resonances.   Their positions can be easily seen from the bottom figure.
The several other peaks are due to two-photon resonances.
The two-photon peaks with large intensities are located at about
$3\omega = 3.0$, 4.8, 6.4, 7.1, and 7.2eV.  The other
resonances have weaker intensities and cannot be seen.

We would like to point out three properties: (1) The magnitude of $\chi^{(3)}$
at $\omega = 0$ is $1.22 \times 10^{-12}$esu and is similar to the
magnitudes in the THG experiments: $4 \times 10^{-12}$esu at $3\omega \simeq
1.6$eV (Meth et al.) and $7 \times 10^{-12}$esu at $3\omega \simeq 3.6$eV
(Meth et al. and Kafafi et al.).
Here, we compare the magnitudes at frequencies far
from the resonances. (2) The value of the THG around the peak at 2.5eV
is of the order of $10^{-11}$esu.  This well explains the magnitude
$2.7 \times 10^{-11}$esu at the peak centered around 2.8eV (Meth et al.).
A larger broadening in our theory would yield better agreement
with the experiment.  (3) The three-photon peaks, at $3\omega
\simeq 6.1$ and 6.3eV, have remarkably large strengths, when compared with
the relative smallness of the one-photon absorption at the same energies
(Fig. 1).  This large enhancement would be due
to the fact that there are two-photon resonances
at $3\omega \simeq 6.1$, 6.4, and 6.5eV, meaning that double resonance
enhancement occurs.  To our knowledge,
experiments over this higher energy region have
not been reported yet.  Verification is urgently needed.

In the above discussion, we have not included the local field correction
effect.  Meth et al. have estimated that the nonlinearity would be
enhanced by a factor of 16.  On the other hand, plasmon excitations
due to Coulomb interactions among $\pi$-electrons might screen the
external field and the strength of the optical response might be suppressed, as
discussed in ref. 8 and the report by Wang et al.\footnote[1]{Y. Wang,
G. F. Bertsch and D. Tom\'{a}nek: preprint.}
The RPA treatment by Wang et al. gave a reduction factor as small as on the
order of 10$^{-2}$.  However, this value seems to be overestimated.
Even though we have not taken into account both effects,
we have obtained magnitudes for THG comparable to those of experiments
in the energy region $3\omega \simle 3$eV.  This fact might indicate that
the enhancement by the local field effect and the reduction due to
the Coulomb screening are mutually balanced in this region.

\section{Conclusions}

We have calculated the dispersion relation of the THG spectrum,
making use of the simple tight-binding model for
$\pi$-electron systems of $\soc$.
We have obtained the magnitudes of the THG, which agree with those
in experiments in the lower energy region from $3\omega = 0$ to
3eV.  For higher energies, where experimental
information is lacking at present,
large enhancement of the THG at $3\omega \sim 6$eV has been obtained.
This is due to the double resonance enhancement by two-photon
and three-photon processes.

In the present paper, the Coulomb interactions among $\pi$-electrons
have not been taken into account.  The shift of the oscillator strength
to higher energy transitions due to the Coulomb effect$^{8)}$
would further enhance the strong THG peaks around $3\omega \sim 6$eV.
Therefore, the Coulomb effect does not change our conclusion of
double resonance enhancement.

The Coulomb interactions might change the relative
ratio of strengths of the peaks, as discussed in one-photon
optical absorption,$^{8)}$  even though the fundamental magnitude
would be determined by the present calculation.  We have not considered
the local field correction also.  The correction factor should be
determined by a microscopic theory with Coulomb interactions.
These problems are left for future studies.

\pagebreak
\begin{flushleft}
{\bf References}
\end{flushleft}

\noindent
1) A. F. Hebard, M. J. Rosseinsky, R. C. Haddon, D. W. Murphy,
S. H. Glarum, T. T. M. Palstra, A. P. Ramirez and A. R. Kortan:
Nature {\bf 350} (1991) 600.\\
2) P. M. Allemand, K. C. Khemani, A. Koch, F. Wudl, K. Holczer, S. Donovan,
G. Gr\"{u}ner and J. D. Thompson: Science {\bf 253} (1991) 301.\\
3) W. J. Blau, H. J. Byrne, D. J. Cardin, T. J. Dennis, J. P. Hare,
H. W. Kroto, R. Taylor and D. R. M. Walton: Phys. Rev. Lett. {\bf 67}
(1991) 1423.\\
4) H. Hoshi, N. Nakamura, Y. Maruyama, T. Nakagawa, S. Suzuki,
H. Shiromaru and Y. Achiba: Jpn. J. Appl. Phys. {\bf 30} (1991) L1397.\\
5) N. Bloembergen: {\sl Nonlinear Optics} (Benjamin, New York, 1965).\\
6) B. J. Orr and J. F. Ward: Mol. Phys. {\bf 20} (1971) 513.\\
7) J. Yu, B. Friedman, P. R. Baldwin and W. P. Su:
Phys. Rev. B {\bf 39} (1989) 12814.\\
8) G. F. Bertsch, A. Bulgac, D. Tom\'{a}nek and Y. Wang:
Phys. Rev. Lett. {\bf 67} (1991) 2690.\\
9) S. Saito and A. Oshiyama: Phys. Rev. Lett. {\bf 66} (1991) 2637.\\
10) G. W. Hayden and E. J. Mele: Phys. Rev. B {\bf 36} (1987) 5010.\\
11) K. Harigaya: J. Phys. Soc. Jpn. {\bf 60} (1991) 4001.\\
12) K. Harigaya: to be published in Phys. Rev. B (1992).\\
13) J. Yu and W. P. Su: Phys. Rev. B {\bf 44} (1991) 13315.\\
14) J. R. Heath, R. F. Curl and R. E. Smalley:
J. Chem. Phys. {\bf 87} (1987) 4236.\\

\pagebreak
\begin{flushleft}
{\bf Figure Captions}
\end{flushleft}

\noindent
Fig. 1. The single-photon absorption spectrum.  The unit of the ordinate
is arbitrary.  In the lower figure, we show energy values of the
dipole allowed and forbidden transitions.

{}~

\noindent
Fig. 2. The third harmonic generation spectra.  Figures (a) and
(b) are the real and imaginary parts, respectively.  In (c),
the absolute value is depicted.
In the bottom figure, we show the energies
of dipole allowed excitations as well as the energies of dipole
forbidden (two-photon allowed) excitations multiplied by 1.5.


\end{document}